\documentclass[aps,prd,twoside,twocolumn,nofootinbib,showpacs,floatfix]{revtex4-1}
\usepackage{amssymb}
\usepackage{amsmath,amssymb}
\usepackage{graphicx,bm}
\usepackage{slashed}
\usepackage{times}
\usepackage{epstopdf}
\usepackage{ulem} 
\usepackage[usenames]{color}
\usepackage{float}
\usepackage{subfigure}
\usepackage{subfigure}
\usepackage{rotating}
\usepackage{color}
\usepackage{multirow}
\usepackage{dcolumn}
\usepackage{overpic}
\usepackage{booktabs}
\usepackage{appendix}
\usepackage{makecell}
\usepackage{diagbox}

\renewcommand\sout{\bgroup \color{red} \ULdepth=-.5ex \ULset}

\usepackage[colorlinks, citecolor=blue,anchorcolor=red,menucolor=red,linkcolor=blue,filecolor=red,runcolor=red,urlcolor=blue,frenchlinks=red]{hyperref}
\begin{document}
\title{$D$-wave dimeson resonance interpretation to the $Z_c/Z_{cs}/Z_b$ states}
\author{Kan Chen}
\affiliation{School of Physics, Northwest University, Xi'an 710127, China}
\affiliation{Shaanxi Key Laboratory for theoretical Physics Frontiers, Xi'an 710127, China}
\affiliation{Institute of Modern Physics, Northwest University, Xi'an 710127, China}
\affiliation{Peng Huanwu Center for Fundamental Theory, Xi'an 710127, China}
\author{Jun-Zhang Wang}\email{wangjzh@cqu.edu.cn}
\affiliation{Department of Physics, Chongqing University, Chongqing 401331, China}
\begin{abstract}
In this work, we construct the $S$-wave, $P$-wave, and $D$-wave $D^{(*)}\bar{D}^{*}_{(s)}$/$B^{(*)}\bar{B}^{*}_{(s)}$ interactions by considering the lowest order contributions from the leading order (LO), next-to-leading order (NLO), and next-to-next-to-next-to-leading order (N$^3$LO) contact terms, respectively. After solving the corresponding Lippmann-Schwinger equation (LSE), we obtain the typical pole trajectory for the $S$-wave/$P$-wave/$D$-wave state. We adjust the $P$-wave low energy constant (LEC) and {accordingly obtain a  $P$-wave $D\bar{D}^*/D^*\bar{D}$} resonance state that may correspond to the observed $G(3900)$, { and its $P$-wave $D^*\bar{D}^*$ and $B^{(*)}\bar{B}^{*}$ partners are also predicted}. Then we proceed to adjust the $D$-wave LECs to looking for the $D$-wave $D^{(*)}\bar{D}^{(*)}_{(s)}$/$B^{(*)}\bar{B}^{(*)}_{(s)}$ resonances. The satisfactory consistencies between our results { and the experimental resonance parameters} of $Z_c(3900)$/$Z_c(4020)$, $Z_b(10610)$/$Z_b(10650)$, and $Z_{cs}(4000)$/$Z_{cs}(4220)$ indicate that the $D$-wave resonance interpretation may provide a unified picture to understand these { isovector charmoniumlike states. Especially, the broad width features of $Z_{cs}(4000)$ and $Z_{cs}(4220)$ can be naturally reproduced by interpreting them as $D$-wave $D\bar{D}_s^*$ and $D^*\bar{D}_s^*$ resonances, respectively.} 
\end{abstract}
\maketitle
\section{Introduction}

In the past decade, many exotic states were observed experimentally \cite{Chen:2016qju,Lebed:2016hpi,Esposito:2016noz,Chen:2016spr,Guo:2017jvc,Olsen:2017bmm,Brambilla:2019esw,Chen:2022asf,Meng:2022ozq,Liu:2024uxn}. Especially the charged charmonium-like $Z_c$ states \cite{BESIII:2013ris,BESIII:2013qmu,BESIII:2015pqw,BESIII:2013ouc,BESIII:2015tix,Belle:2013yex,Xiao:2013iha} ($Z_c(3900)$ and $Z_c(4020)$), bottomonium-like $Z_b$ states \cite{Belle:2011aa,Belle:2015upu} ($Z_b(10610)$ and $Z_b(10650)$), and $Z_{cs}$ states \cite{LHCb:2023hxg,LHCb:2021uow} ($Z_{cs}(4000)$ and $Z_{cs}(4220)$). These states carry exotic flavor quantum numbers and are good candidates of either hadronic molecules composed of a pair of heavy mesons or hidden-charm tetraquark states.

In the last year, the BESIII collaboration observed a structure around 3.9 GeV \cite{BESIII:2024ths} (we refer to this structure as $G(3900)$ in the following discussion \cite{BaBar:2006qlj,BaBar:2008drv,Belle:2007qxm}) from the measurement of Born cross sections for the $e^+e^-\rightarrow D\bar{D}$ process. Its mass and width (see the
supplemental material of Ref. \cite{BESIII:2024ths}) were obtained as $3872.2\pm14.2\pm3.0$ MeV and $179.7\pm14.1\pm7.0$ MeV, respectively.
The $G(3900)$ was investigated by considering the { coupled channel interference effects} \cite{Eichten:1979ms,Husken:2024hmi}, {threshold cusp effect \cite{Zhang:2009gy}}, final state interaction \cite{Salnikov:2024wah}, while the prevailing interpretation of this state is a $P$-wave $D\bar{D}^*$ resonance \cite{Lin:2024qcq,Nakamura:2023obk,Huang:2025rvj,BESIII:2025wlf,Chen:2025gxe,Du:2016qcr,Ye:2025ywy,Liu:2025sjz}.

The $G(3900)$, together with the aforementioned $Z_c$/$Z_{b}$/$Z_{cs}$ are all close to one pair of heavy mesons. Thus, the molecular configuration becomes a nature interpretation to these states. Especially, the $Z_{c}(3900)$ can be interpreted as an $S$-wave $D\bar{D}^*$ virtual state within the framework of one-boson-exchange model (OBE) \cite{Lin:2024qcq,Yu:2024sqv,He:2017lhy} and {effective field theory }\cite{Albaladejo:2015lob,Du:2022jjv,Gong:2016hlt}.
The masses of $Z_c(3900)$ and $G(3900)$ are very close to each other, this fact can be understood easily in the molecular picture since they are bounded by the residual strong interaction between the $D$ and $\bar{D}^*$ mesons, the excitation of angular momentum leads to only several MeVs of mass shift from the $D\bar{D}^*$ threshold.

{Although the $Z_c(3900)$ could be interpreted as an $S$-wave virtual state, however, its mass determined from different groups all lie above the $D\bar{D}^*$ threshold by several MeVs \cite{ParticleDataGroup:2024cfk}. Together with its non-negligible decay width, thus, instead of a virtual state, the $Z_c(3900)$ is more like a resonance state. This difficulty also arises in the resonance parameters of observed $Z_c(4020)$/$Z_{b}(10610)$/$Z_b(10650)$ states.}

The $Z_c(3900)$/$Z_c(4020)$, $Z_b(10610)$/$Z_b(10650)$, and $Z_{cs}(4000)$/$Z_{cs}(4220)$ states, with $J^P=1^+$ \cite{ParticleDataGroup:2024cfk}, { were widely considered to be the candidate of the ${}^{3}S_1$ states of $D\bar{D}^*$/$D^*\bar{D}^*$, $B\bar{B}^*$/$B^*\bar{B}^*$, and $D\bar{D}_s^*$/$D^*\bar{D}_s^*$ \cite{Chen:2016qju,Lebed:2016hpi,Esposito:2016noz,Chen:2016spr,Guo:2017jvc,Olsen:2017bmm,Brambilla:2019esw,Chen:2022asf,Meng:2022ozq,Liu:2024uxn}, respectively.} Besides, the $J^P=1^+$ can also correspond to the ${}^3D_1$ states. To our knowledge, this possibility has not been discussed in the literatures.

Due to the repulsive effect of centrifugal barrier potential, the states with angular momentum $L>0$ can hardly form bound states, but it is still possible to form resonances \cite{Taylor}, which is exactly the case for the observed $G(3900)$. The closeness between the measured masses of $Z_c(3900)$ and $G(3900)$ leads us to conjecture that if the ${}^{3}D_1$ $D\bar{D}^*$ state could form resonance state, then the ${}^3S_1$ $D\bar{D}^*$ state and ${}^3D_1$ $D\bar{D}^*$ state may lie very close to each other, i.e., the excitation of angular momentum with $L=2$ only leads to a few MeVs mass shift with respect to the $D\bar{D}^*$ threshold, so that the $S$-wave and $D$-wave states may coexist in a narrow energy region { and may simultaneously contribute into these observed charged near-threshold structures}. This might be a reasonable solution to clarify the puzzling resonance parameters of $Z_c(3900)$/$Z_c(4020)$ and $Z_b(10610)$/$Z_b(10650)$ states.

In this work, we construct an effective field theory by considering the heavy quark symmetry (HQS) and SU(3) flavor symmetry (SU(3)$_{\text{f}}$). We approximate the $S$-wave, $P$-wave, and $D$-wave effective potentials by considering the lowest order contributions from the LO, NLO, and N$^3$LO contact terms \cite{Epelbaum:2004fk}, respectively. With the obtained $S$-wave, $P$-wave, and $D$-wave potentials, we solve the corresponding Lippmann-Schwinger equation (LSE). By adjusting the LECs introduced from the $S$-wave, $P$-wave, and $D$-wave contact potentials, we present the general pole trajectory behaviors of the $S$-wave, $P$-wave, and $D$-wave states. Then we attempt to reproduce the experimentally observed $G(3900)$/$Z_c$/$Z_b$/$Z_{cs}$ resonance states and predict their possible heavy quark spin symmetry (HQSS) partners and heavy quark flavor symmetry (HQFS) partners.

This work is organised as follows. In Sec. \ref{framework}, we present our framework, including the constructions of $S$-wave, $P$-wave, and $D$-wave contact potentials, analytic solutions to the LSEs from two different form factors. In Sec. \ref{Pole trajectories} we present the typical pole trajectories for the $S$-wave, $P$-wave, and $D$-wave states. The approximation we adopted to fix the LECs of $D^{(*)}\bar{D}_{(s)}^*$ and $B^{(*)}\bar{B}_{(s)}^*$ states is discussed in \ref{approximation}. Then we present our numerical results and discussions on the $P$-wave and $D$-wave resonance states in Sec. \ref{results}. Sec. \ref{summary} is devoted to a summary.

\section{Framework}\label{framework}
In this section, we take the $D^{(*)}\bar{D}^{*}$ system as an example to introduce the contact terms describing the interactions of the $S$-wave. $P$-wave, and $D$-wave states. The obtained contact potentials can be directly generalized to the $D^{(*)}\bar{D}^{*}_s$ and $B^{(*)}\bar{B}^{*}_{(s)}$ systems. 
\subsection{Contact potentials}\label{contact}

In the heavy quark limit $m_Q\rightarrow \infty$, the interactions between $D^{(*)}$ and $\bar{D}^{*}$ mesons do not depend on the flavor operators and spin operators from their heavy quarks. Consequently, in this limit, the contact terms depend on the flavor operators $\bm{F}_1$ for light quark and $\bm{F}_2$ for light antiquark in each meson, spin operators $\bm{\sigma}_1$ for light quark and $\bm{\sigma}_2$ for light antiquark in each meson, and the initial and final meson momenta $\bm{p}$ and $\bm{p}^\prime$. Thus, we have
\begin{eqnarray}
V_{\text{Cont}}&=&V\left(\bm{F}_1,\bm{F}_2,\bm{\sigma}_1,\bm{\sigma}_2,\bm{p},\bm{p}^\prime\right).
\end{eqnarray}
For the $D^{(*)}\bar{D}^*$ system, in the SU(3)$_{\text{f}}$ limit, the channels with different isospins are related via the scalar product
\begin{eqnarray}
\bm{F}_1\cdot\bm{F}_2=\sum_{i=1}^8\lambda_1^i\cdot\left(-\lambda_2^{i*}\right),
\end{eqnarray}
where $\lambda^i_{1}$ and $\lambda_2^{i*}$ are the Gell-mann matrix and its conjugate matrix, respectively. We can extract this scalar product out
\begin{eqnarray}
V_{\text{Cont}}&=&\left(\bm{F}_1\cdot\bm{F}_2\right)\tilde{V}\left(\bm{\sigma}_1,\bm{\sigma}_2,\bm{p},\bm{p}^\prime\right).
\end{eqnarray}
By considering the parity invariance and time reversal invariance, the explicit form of the $\tilde{V}\left(\bm{\sigma}_1,\bm{\sigma}_2,\bm{p},\bm{p}^\prime\right)$ can be constructed by analogy with the contact terms of the two-nucleon potential \cite{Epelbaum:2004fk}. Here, since the lowest order contributions to the $D$-wave potentials are from the N$^3$LO contact terms. Thus, up to the N$^3$LO, the contact terms can be written as
\begin{eqnarray}
V_{\text{Cont}}&=&\left(\bm{F}_1\cdot\bm{F}_2\right)\tilde{V}\left(\bm{\sigma}_1,\bm{\sigma}_2,\bm{p},\bm{p}^\prime\right)\nonumber\\
&=&\left(\bm{F}_1\cdot\bm{F}_2\right)\left(V_{\text{Cont}}^{(0)}+V_{\text{Cont}}^{(2)}+V_{\text{Cont}}^{(4)}\right),\label{VTotal}\\
V_{\text{Cont}}^{(0)}&=&C_S+C_T\bm{\sigma}_1\cdot\bm{\sigma}_2,\label{V0}\\
V_{\text{Cont}}^{(2)}&=&C_1\bm{q}^2+C_2\bm{k}^2+\sum_{i=3}^{7}C_i\mathcal{O}_i,\label{V2}\\
V_{\text{Cont}}^{(4)}&=&D_1\bm{q}^4+D_2\bm{k}^{4}+D_3\bm{q}^2\bm{k}^2+D_4\left(\bm{q}\times\bm{k}\right)^2\nonumber\\
&&+\sum_{j=5}^{15}D_j\mathcal{O}^\prime_{j},\label{V4}
\end{eqnarray}
with { the transferred momentum $\bm{q}=\bm{p}^\prime-\bm{p}$ and the average momentum $\bm{k}=(\bm{p}+\bm{p}^\prime)/2$}. ($C_S$, $C_T$), $C_i$ ($i=1,...,7$), and $D_j$ ($j=1,...,15$) are unknown LECs from the LO, NLO, and N$^3$LO contact terms, respectively. The explicit forms of the operators $\mathcal{O}_i$ ($i=3,...,7$) and $\mathcal{O}_j^\prime$ ($j=5.,,,.15$) in Eqs. (\ref{V2}-\ref{V4}) can be found in Ref. \cite{Epelbaum:2004fk}.

We label the total isospin, total spin, angular momentum, total angular momentum, and eigenvalue of $C$-parity of a $D^{(*)}\bar{D}^*$ state with $I$, $s$, $L$, $J$, and $c$, respectively. With the contact terms in Eqs. (\ref{VTotal}-\ref{V4}), the effective potentials of $|^s_I [D^{(*)}\bar{D}^*]^L_J,c\rangle$ states with ${}^{2s+1}L_J$ ($L\equiv S$, $P$, and $D$), isospin $I$ and parity $c$ can be described by the following matrix elements
\begin{widetext}
\begin{eqnarray}
V_{\left|^{s}_{I}[D^{(*)}\bar{D}^*]^S_J,c\right\rangle}
&=&
\left\langle ^s_I[D^{(*)}\bar{D}^*]^S_J,c\right|V_{\text{Cont}}\left| ^s_I[D^{(*)}\bar{D}^*]^S_J,c\right\rangle\nonumber\\
&=&\tilde{C}_{\left|^s_I[D^{(*)}\bar{D}^*]^S_J,c\right\rangle}+C_{\left|^s_I[D^{(*)}\bar{D}^*]^S_J,c\right\rangle}\left(p^2+p^{\prime2}\right)\nonumber\\
&&+D^1_{\left|^s_I[D^{(*)}\bar{D}^*]^S_J,c\right\rangle}p^2p^{\prime2}+D^2_{\left|^s_I[D^{(*)}\bar{D}^*]^S_J,c\right\rangle}\left(p^4+p^{\prime4}\right),\label{STotal}\\
V_{\left|^{s}_{I}[D^{(*)}\bar{D}^*]^P_J,c\right\rangle}
&=&\left\langle ^{s}_{I}[D^{(*)}\bar{D}^*]^P_J,c\right|V_{\text{Cont}}\left| ^{s}_{I}[D^{(*)}\bar{D}^*]^P_J,c\right\rangle\nonumber\\
&=&C_{\left|^{s}_{I}[D^{(*)}\bar{D}^*]^P_J,c\right\rangle}pp^\prime+D_{\left|^{s}_{I}[D^{(*)}\bar{D}^*]^P_J,c\right\rangle}pp^\prime\left(p^2+p^{\prime2}\right),\label{PTotal}\\
V_{\left|^{s}_{I}[D^{(*)}\bar{D}^*]^D_J,c\right\rangle}
&=&\left\langle ^{s}_{I}[D^{(*)}\bar{D}^*]^D_J,c\right|V_{\text{Cont}}\left| ^{s}_{I}[D^{(*)}\bar{D}^*]^D_J,c\right\rangle\nonumber\\
&=&D_{\left|^{s}_{I}[D^{(*)}\bar{D}^*]^D_J,c\right\rangle}p^2p^{\prime2},\label{DTotal}
\end{eqnarray}
\end{widetext}
with $p=|\vec{p}|$ and $p^\prime=|\vec{p}^\prime|$. Here, $\tilde{C}_{|^s_I[D^{(*)}\bar{D}^*]_J^S,c\rangle}$ is the linear combination of the LO LECs $C_S$ and $C_T$, $C_{|^s_I[D^{(*)}\bar{D}^*]_J^S,c\rangle}$ and $C_{|^s_I[D^{(*)}\bar{D}^*]_J^P,c\rangle}$ are the linear combinations of the NLO LECs $C_i$ ($i=1,...,7$). Similarly,
$D^1_{|^s_I[D^{(*)}\bar{D}^*]_J^S,c\rangle}$, $D^2_{|^s_I[D^{(*)}\bar{D}^*]_J^S,c\rangle}$, $D_{|^s_I[D^{(*)}\bar{D}^*]_J^P,c\rangle}$, and $D_{|^s_I[D^{(*)}\bar{D}^*]_J^D,c}\rangle$ are the linear combinations of the N$^{3}$LO LECs $D_j$ ($j=1,...,15$). { The expressions of these parameters in Eqs. (\ref{STotal}-\ref{DTotal}) can be explicitly derived by performing the partial-wave expansion \cite{Golak:2009ri}. Since we are mainly interested in the channels that have the same quantum numbers to that of the $Z_c(3900)$ and $G(3900)$ states, so we do not need the explicit forms of these parameters. For conciseness, their complete expressions in Eqs. (\ref{STotal}-\ref{DTotal}) will not be presented here.}

Note that the operators $\bm{\sigma}_1$ and $\bm{\sigma}_2$ in Eqs. (\ref{V0}-\ref{V4}) only act on the spins of light quark and light antiquark, respectively, thus, up to the N$^{3}$LO, for the $S$-wave $D^{(*)}\bar{D}^{(*)}$ states, the HQSS requires the following two relations
\begin{eqnarray}
V_{\left|^1_I[D\bar{D}^*]_1^S,-\right\rangle}&=&V_{\left|^1_I[D^*\bar{D}^*]_1^S,-\right\rangle},\label{SHQSS1}\\
V_{\left|^1_I[D\bar{D}^*]_1^S,+\right\rangle}&=&V_{\left|^2_I[D^*\bar{D}^*]_2^S,+\right\rangle}.\label{SHQSS2}
\end{eqnarray}
For the $P$-wave $D^{(*)}\bar{D}^{(*)}$ states, we have
\begin{eqnarray}
V_{\left|^1_I[D\bar{D}^*]_0^P,+\right\rangle}&=&V_{\left|^1_I[D^*\bar{D}^*]_0^P,+\right\rangle},\label{PHQSS1}\\
V_{\left|^1_I[D\bar{D}^*]_1^P,-\right\rangle}&=&V_{\left|^1_I[D^*\bar{D}^*]_1^P,-\right\rangle},\label{PHQSS2}\\
V_{\left|^1_I[D\bar{D}^*]_2^P,+\right\rangle}&=&V_{\left|^1_I[D^*\bar{D}^*]_2^P,+\right\rangle},\label{PHQSS3}
\end{eqnarray}
and
\begin{eqnarray}
V_{\left|^1_I[D\bar{D}^*]_1^D,-\right\rangle}&=&V_{\left|^1_I[D^*\bar{D}^*]_1^D,-\right\rangle},\label{DHQSS1}\\
V_{\left|^1_I[D\bar{D}^*]_2^D,+\right\rangle}&=&V_{\left|^1_I[D^*\bar{D}^*]_2^D,+\right\rangle},\label{DHQSS2}\\
V_{\left|^1_I[D\bar{D}^*]_3^D,-\right\rangle}&=&V_{\left|^1_I[D^*\bar{D}^*]_3^D,-\right\rangle}\label{DHQSS3}
\end{eqnarray}
for the $D$-wave $D^{(*)}\bar{D}^{(*)}$ states.

The relations in Eqs. (\ref{SHQSS1}-\ref{DHQSS3}) can be directly applied to the $B^{(*)}\bar{B}^{(*)}$ system according to the HQFS. Besides, by assigning the $D^{(*)}\bar{D}_{s}^{(*)}$ state as the corresponding SU(3)$_\text{f}$ partner of the $D^{(*)}\bar{D}^{(*)}$ state, the above relations can also be directly applied to the $D^{(*)}\bar{D}_{s}^{(*)}$ system.

We introduce 24 LECs in Eqs. (\ref{V0}-\ref{V4}), $C_S$ and $C_T$ from the LO contact terms, 7 $C_i$s from the NLO contact terms, and 15 $D_j$s from the N$^{3}$LO contact terms. To reduce the number of LECs, we further assume that the emergences of $S$-wave/$P$-wave/$D$-wave bound/virtual/resonance states are dominated by the lowest order contributions from the $S$-wave/$P$-wave/$D$-wave contact terms. Specifically, for the $S$-wave contact potential in Eq. (\ref{STotal}), we only consider the constant term and omit all the momentum-dependent terms, for the $P$-wave contact potential in Eq. (\ref{PTotal}), we drop the $pp^\prime(p^2+p^{\prime 2})$ term.

Besides, since we are mainly interested in the $I(J^{PC})=1(1^{+-})$ channel for the $S$-wave or $D$-wave $D\bar{D}^{*}$ state, and the $I(J^{PC})=0(1^{--})$ channel for the $P$-wave $D\bar{D}^{*}$ state. Thus, the contact terms for the $I=1$ $S$-wave and $D$-wave $D\bar{D}^*$ states with $J^{PC}= 1^{+-}$, and the contact term for the $I=0$ $P$-wave $D\bar{D}^*$ states with $J^{PC}= 1^{--}$ can be {rewritten} as
\begin{eqnarray}
V_{\left|^1_1[D\bar{D}^{*}]^S_1,-\right\rangle}&=&g_0^{D\bar{D}^*},\label{Sg0}\\
V_{\left|^1_1[D\bar{D}^{*}]^D_1,-\right\rangle}&=&g_2^{D\bar{D}^*}p^2p^{\prime 2},\label{Dg2}\\
V_{\left|^1_0[D\bar{D}^{*}]^P_1,-\right\rangle}&=&g^{D\bar{D}^*}_1pp^{\prime}.\label{Pg1}
\end{eqnarray}
Here, we simplify our notations by replacing the redefined LECs $\tilde{C}_{|^1_1[D\bar{D}^{*}]^S_1,-\rangle}$, $C_{|^1_0[D\bar{D}^{*}]^P_1,-\rangle}$, and $D_{|^1_1[D\bar{D}^{*}]^D_1,-\rangle}$ with $g^{D\bar{D}^*}_1$, $g^{D\bar{D}^*}_2$, and $g^{D\bar{D}^*}_3$, respectively. For the HQSS and HQFS partners of the $|^1_1[D\bar{D}^{*}]^S_1,-\rangle$, $|^1_0[D\bar{D}^{*}]^P_1,-\rangle$, and $|^1_1[D\bar{D}^{*}]^D_1,-\rangle$ states, in Sec. \ref{approximation}, we will adopt further approximations to estimate their LECs.    

\subsection{Lippmann-Schwinger equation (LSE)}\label{LSE}
With the obtained contact potential for the $|^s_I[D^{(*)}\bar{D}^*]_J^L,c\rangle$ state, we need to further suppress the contributions from higher momenta by introducing a regulator $F_\Lambda(p)$
\begin{eqnarray}
V^{\Lambda}_{\left|^s_I[D^{(*)}\bar{D}^*]_J^L,c\right\rangle}(p,p^\prime)&=&
V_{\left|^s_I[D^{(*)}\bar{D}^{*}]_J^L,c\right\rangle}F_{\Lambda}(p^\prime)F_{\Lambda}(p),
\end{eqnarray}
with $L\equiv S$, $P$, $D$. Since the form factor $F_{\Lambda}(p)$ introduces extra momentum-dependent effect to the effective potential $V^{\Lambda}_{|^s_I[D^{(*)}\bar{D}^*]_J^L,c\rangle}(p,p^\prime)$, our numerical results will depend on the explicit form of $F_{\Lambda}(p)$. To check the form factor dependences of our results, we introduce the following two types of form factors , i.e., the Gaussian form
\begin{eqnarray}
F_{\Lambda}(p)=\text{exp}(-\frac{p^2}{\Lambda^2}),\label{FF1}
\end{eqnarray}
and the hard cutoff form
\begin{eqnarray}
F_{\Lambda}(p)=\theta\left(\Lambda-p\right)\left(1-\frac{p^2}{\Lambda^2}\right).\label{FF2}
\end{eqnarray}
We refer the scenarios in Eqs. (\ref{FF1}) and (\ref{FF2}) as Model I and Model II, respectively. The Gaussian type form factor in model I is widely adopted to study the two-nucleon interactions \cite{Epelbaum:2008ga,Machleidt:2011zz} and the hadronic molecules \cite{Wang:2019ato,Meng:2019ilv,Du:2019pij,Du:2021zzh}. In model II \cite{Gasparyan:2025dpj,Epelbaum:2017byx,Meng:2021rdg}, the step function $\theta(\Lambda-p)$ is adopted to cut the contribution from $\Lambda>p$ region, since we will also study the $P$-wave and $D$-wave interactions, we further multiple a factor $(1-p^2/\Lambda^2)$ to obtain a better description to the states with higher partial waves.

The cutoff $\Lambda$ in the form factor of model I is fixed at 0.5 GeV. { We find that it will not change our results qualitatively by varying the cutoff value in a reasonable region, and will also not change the pole trajectory behaviors of the $S$-wave, $P$-wave, and $D$-wave bound/virtual/resonance states.} These conclusions also hold for the results obtained from the form factor in model II, so in model II, we also fix the cutoff $\Lambda$ at 0.5 GeV.

The general Lippmann-Schwinger equation (LSE) reads
\begin{eqnarray}
T\left(\bm{p},\bm{p^\prime}\right)&=&V\left(\bm{p},\bm{p}^\prime\right)+\int\frac{d^3\bm{q}}{(2\pi)^3}\frac{V\left(\bm{p},\bm{q}\right)T\left(\bm{q},\bm{p}^\prime\right)}{E-\frac{\bm{q}^2}{2\mu}+i\epsilon},
\end{eqnarray}
where $\mu$ is the reduced mass and $E$ is the energy with respect to the threshold. With the separable potentials for $L=0$, 1, and 2 in model I and II, the corresponding LSE can be solved analytically (more details can be found in Ref. \cite{Ye:2025ywy} for model I and Ref. \cite{Meng:2021rdg} for model II).

The pole solutions of the bound/virtual/resonance states satisfy the following expression
\begin{eqnarray}
1-g_{L} \mu G_{2L}(k)=0,\label{Pole solution}
\end{eqnarray}
with $L=0$, $1$, $2$ for the $S$-wave, $P$-wave, and $D$-wave potentials, respectively. $k$ is defined by $k=\sqrt{2\mu E}$.

In model I, the $S$-wave two-point function $G_{0}(k)$ can be calculated as
\begin{eqnarray}
G_0\left(k\right)&=&\int\frac{d^3\bm{q}}{\left(2\pi\right)^3}\frac{\text{exp}\left(-\frac{2q^2}{\Lambda^2}\right)}{\frac{k^2}{2}-\frac{q^2}{2}+i\epsilon}\\
&=&-\frac{\Lambda}{\left(2\pi\right)^{\frac{3}{2}}}+\frac{k}{2\pi}\text{exp}\left(-\frac{2k^2}{\Lambda^2}\right)\left[\text{erfi}\left(\frac{\sqrt{2}k}{\Lambda}\right)-i\right],\label{MIG0}\nonumber\\
\end{eqnarray}
where
\begin{eqnarray}
\text{erfi}\left(\frac{\sqrt{2}k}{\Lambda}\right)&=&\frac{2}{\sqrt{\pi}}\int_0^{\frac{\sqrt{2}k}{\Lambda}}dt\text{exp}(t^2)
\end{eqnarray}
is the error function, and
\begin{eqnarray}
G_{m}\left(k\right)=\frac{\Lambda^3}{4}\frac{\partial G_{m-2}\left(k\right)}{\partial \Lambda},\label{MIGm}
\end{eqnarray}
with $m=2$, $4$. We can use Eq. (\ref{MIG0}) and iterate Eq. (\ref{MIGm}) to obtain the explicit forms of $P$-wave two-point function $G_2(k)$ and $D$-wave two-point function $G_4(k)$.

In model II, the $G_{2L}(k)$ function is given by
\begin{eqnarray}
G_{2L}=\mathcal{G}_{2L}-\frac{2}{\Lambda^2}\mathcal{G}_{2L+2}+\frac{1}{\Lambda^4}\mathcal{G}_{2L+4},
\end{eqnarray}
where $\mathcal{G}_0(k)$ can be calculated as
\begin{eqnarray}
\mathcal{G}_0&=&\int\frac{d^3\bm{q}}{\left(2\pi\right)^3}\frac{\left[\theta\left(\Lambda-q\right)\right]^2}{\frac{k^2}{2}-\frac{q^2}{2}+i\epsilon}\nonumber\\
&=&\frac{1}{\pi^2}\left[k\text{tanh}^{-1}\left(\frac{k}{\Lambda}\right)-\Lambda-i\frac{\pi}{2}k\right],\label{MIIG0}
\end{eqnarray}
and
\begin{eqnarray}
\mathcal{G}_{m}\left(k\right)=k^2\mathcal{G}_{m-2}-\frac{1}{\pi^2}\frac{\Lambda^{m+1}}{m+1},\label{MIIGm}
\end{eqnarray}
with $m=2$, 4, 6, 8. We can use the result in Eq. (\ref{MIIG0}) and iterate Eq. (\ref{MIIGm}) to obtain the explicit forms of $\mathcal{G}_2(k)$, $\mathcal{G}_4(k)$, $\mathcal{G}_6(k)$, and $\mathcal{G}_8(k)$.

Note that in Eq. (\ref{Pole solution}), we extract the reduced mass $\mu$ out, and the two-point function $G_{2L}(k)$ is only a function of variables $k$ and $\Lambda$, we use this definition for later convenience.
\section{Typical pole trajectories of the $S$-wave, $P$-wave, and $D$-wave states}\label{Pole trajectories}
To have attractive forces, we assume that the LECs $g_1^{D\bar{D}^*}$, $g_2^{D\bar{D}^*}$, and $g_3^{D\bar{D}^*}$ for the $|^1_1[D\bar{D}^*]_1^S,-\rangle$, $|^1_0[D\bar{D}^*]_1^P,-\rangle$, and $|^1_1[D\bar{D}^*]_1^D,-\rangle$ channels are all negative.

For the bound/virtual state, its pole position can be solved by the equation
\begin{eqnarray}
1-g^{D\bar{D}^*}_L\mu_{D\bar{D}^*}G_{2L}\left(k\right)=0,\label{BVsolution}
\end{eqnarray}
with $L=0$, 1, 2, and $\mu_{D\bar{D}^*}$ is the reduced mass of the $D\bar{D}^*$ system. The LEC $g_L^{D\bar{D}^*}$ is solely constrained by the mass of bound/virtual state.

But for the resonance (anti-resonance) solution with both non-zero real part and imaginary part at $k$/$E$ plane, Eq. (\ref{BVsolution}) should be divided into two independent equations
\begin{eqnarray}
\text{Re}\left(1-g^{D\bar{D}^*}_L \mu_{D\bar{D}^*} G_{2L}\left(k\right)\right)&=&0,\label{ReRsolution}\\
\text{Im}\left(1-g^{D\bar{D}^*}_L \mu_{D\bar{D}^*} G_{2L}\left(k\right)\right)&=&0,\label{ImRsolution}
\end{eqnarray}
with $L=1$, 2 (the $S$-wave potential with $L=0$ in Eq. (\ref{Sg0}) can not form resonance). By selecting the LEC $g^{D\bar{D}^*}_L$ at resonance region, we can simultaneously solve the mass and width of resonance state from Eqs. (\ref{ReRsolution}-\ref{ImRsolution}). Thus, the LEC $g_L^{D\bar{D}^*}$ for the resonance state can be constrained by its mass and width simultaneously.

Here, we use the form factor in model I as an example to present the typical pole trajectories for the $S$-wave, $P$-wave, and $D$-wave states. For an illustrative discussion, in Eqs. (\ref{BVsolution}-\ref{ImRsolution}) we firstly set the reduced mass $\mu_{D\bar{D}^*}=1$ GeV (close to the reduced mass of $D^{(*)}\bar{D}^*$ system), then we adjust the LECs $g_1^{D\bar{D}^*}$, $g_2^{D\bar{D}^*}$, and $g_3^{D\bar{D}^*}$ {to sufficiently small negative values so that we can obtain strong enough attractions to form bound states for the $S$-wave, $P$-wave, and $D$-wave interactions, respectively.}

\begin{figure*}[htbp]
    \centering
    \includegraphics[width=1.0\linewidth]{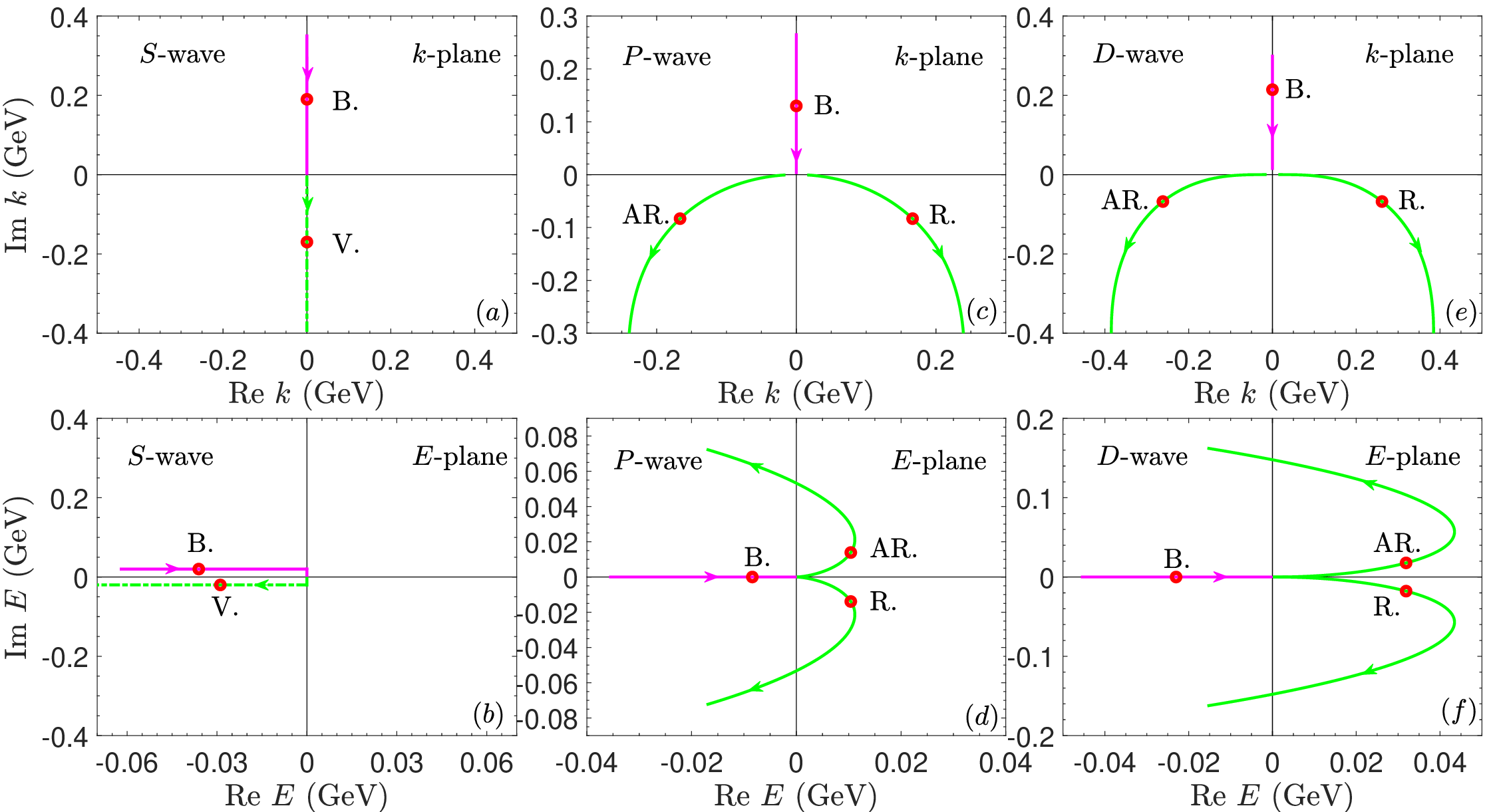}
    \caption{The $S$-wave pole trajectories at $k$-plane (a) and $E$-plane (b), $P$-wave pole trajectories at $k$-plane (c) and $E$-plane (d), and $D$-wave pole trajectories at $k$-plane (e) and $E$-plane (f). We set the reduced masses of $S$-wave, $P$-wave, and $D$-wave two meson states at $\mu=1$ GeV. We use the notations B., V., R., and AR. to label the bound, virtual, resonance, and anti-resonance states, respectively. In sub-figure (b), the poles on the negative real axis are slightly shifted for transparency.}
    \label{spd}
\end{figure*}

In Figs. \ref{spd} (a) and (b), we plot the $S$-wave pole trajectories with $\mu_{D\bar{D}^*}=1$ GeV at $k$-plane and $E$-plane, respectively. The arrows illustrate the direction of decreasing the attractive force (the direction of increasing the LEC $g_0^{D\bar{D}^*}$). As seen from Fig. \ref{spd} (b), when the attractive force is strong enough, the $D\bar{D}^{*}$ forms a bound state, then as the attractive force decreases, the bound state pole moves forward to the threshold alone the negative real axis, after reaching the $D\bar{D}^*$ threshold, it becomes a virtual state and moves backward alone the negative real axis.

The $P$-wave pole trajectories at $k$-plane and $E$-plane are illustrated in Figs. \ref{spd} (c) and (d), respectively. As shown in Fig. \ref{spd} (d), with a strong attractive force, the $P$-wave potential will also lead to bound state. As we decrease the attractive force (increase the LEC $g_1^{D\bar{D}^*}$), the bound state pole moves forward to the threshold and then away from the threshold and finally becomes a resonances/anti-resonance state.

The $D$-wave pole trajectories at $k$-plane and $E$-plane are illustrated in Figs. \ref{spd} (e) and (f), respectively. The pole trajectory behaviors { obtained with the} $D$-wave contact potential at $k$-plane and $E$-plane are very similar to that { obtained with} the $P$-wave contact potential, i.e., the $D$-wave bound state moves forward to the threshold and then becomes a resonance/antiresonance when the interaction changes from a very strong attraction to a weak one. But with a higher angular momentum, the $D$-wave resonance may have a larger mass and broader width compare to that of the $P$-wave resonance if we adjust the LECs $g_1^{D\bar{D}^*}$ and $g_2^{D\bar{D}^*}$, as illustrated in Figs. \ref{spd} (d) and (f), respectively.

We find that the $S$-wave/$P$-wave/$D$-wave pole behaviors obtained from the model II are very similar to that of the model I, therefore, we do not further illustrate the typcial $S$-wave/$P$-wave/$D$-wave pole behaviors obtained from the model II.
\section{Approximation to the LECs in the $D^{(*)}\bar{D}_{(s)}^{*}$ and $B^{(*)}\bar{B}_{(s)}^*$ systems}\label{approximation}
With the limited experimental inputs, we want to further relate the LECs ($g_0^{D\bar{D}^*}$ $g_1^{D\bar{D}^*}$, $g_2^{D\bar{D}^*}$) of the $D\bar{D}^*$ system to the LECs ($g_0^{D^*\bar{D}^*}$, $g_1^{D^*\bar{D}^*}$, $g_2^{D^*\bar{D}^*}$) of its HQSS partner $D^*\bar{D}^*$ system, to the LECs ($g_0^{B\bar{B}^*}$, $g_1^{B\bar{B}^*}$, $g_2^{B\bar{B}^*}$) of its HQFS partner $B\bar{B}^*$ system, and to the LECs ($g_0^{B^*\bar{B}^*}$, $g_1^{B^*\bar{B}^*}$, $g_2^{B^*\bar{B}^*}$) of its HQS partner $B^*\bar{B}^*$ system.

Similarly, the LECs ($g_0^{D\bar{D}_s^*}$ $g_1^{D\bar{D}_s^*}$, $g_2^{D\bar{D}_s^*}$) of the $D\bar{D}_s^*$ system should be related to LECs ($g_0^{D^*\bar{D}_s^*}$, $g_1^{D^*\bar{D}_s^*}$, $g_2^{D^*\bar{D}_s^*}$) of its HQSS partner $D^*\bar{D}_s^*$ system, to the LECs ($g_0^{B\bar{B}_s^*}$, $g_1^{B\bar{B}_s^*}$, $g_2^{B\bar{B}_s^*}$) of its HQFS partner $B\bar{B}_s^*$ system, and to the LECs ($g_0^{B^*\bar{B}_s^*}$, $g_1^{B^*\bar{B}_s^*}$, $g_2^{B^*\bar{B}_s^*}$) of its HQS partner $B^*\bar{B}_s^*$ system.

The magnitudes of LECs in the charm and beauty sectors are discussed in some literatures \cite{AlFiky:2005jd,Baru:2018qkb,Nieves:2011zz}. In Ref. \cite{AlFiky:2005jd}, a general dimensional analysis suggested that the LO LECs are scaled as $1/M$, where $M$ is proportional to the heavy meson mass. In Ref. \cite{Baru:2018qkb}, the authors suggested that a renormalisable EFT requires that the LO
LECs scale as at least $1/M$ in the limit $M\rightarrow \infty$. However, they also concluded that the relations of LECs between different heavy quark sectors may leads to predictions with uncontrolled uncertainties. In addition, the assumption that the contact interaction between the $D$ ($B$) and $\bar{D}^*$ ($\bar{B}^*$) mesons are saturated by the $D$ ($B$) meson weak decay constant $f_D$ ($f_B$) was made in Ref. \cite{Nieves:2011zz} to estimate the LO LEC of $D\bar{D}^*$ and $B\bar{B}^*$ systems.

We take the $D\bar{D}^*$ and $B\bar{B}^*$ systems as examples to present the scheme to estimate the introduced LECs. For the considered $S$-wave $^1_1[D\bar{D}^*/B\bar{B}^*]_1^S,-\rangle$, $P$-wave $|^1_0[D\bar{D}^*/B\bar{B}^*]_1^P,-\rangle$, and $D$-wave $|^1_1[D\bar{D}^*/B\bar{B}^*]_1^D,-\rangle$ states, they satisfy the following pole solution equations
\begin{eqnarray}
1-g_{L}^{D\bar{D}^*}\mu_{D\bar{D}^*} G_{2L}\left(k\right)&=&0,\label{DDstar}\\
1-g_{L}^{B\bar{B}^*}\mu_{B\bar{B}^*} G_{2L}\left(k\right)&=&0,\label{BBstar}
\end{eqnarray}
{ where $g_L^{D\bar{D}^*}$ ($g_{L}^{B\bar{B}^*}$) and $\mu_{D\bar{D}^*}$ ($\mu_{B\bar{B}^*}$) are} the LEC and reduced mass for the $D\bar{D}^*$ ($B\bar{B}^*$) system, respectively. As discussed in Sec. \ref{LSE}, we define $G_{2L}(k)$ as a function of $k$ and $\Lambda$, thus the two-point function $G_{2L}(k)$ do not depend on specific system. The only difference between the pole solution Eqs. (\ref{DDstar}) and (\ref{BBstar}) for the $D\bar{D}^*$ and $B\bar{B}^*$ states are from the factors $g_{L}^{D\bar{D}^*}\mu_{D\bar{D}^*}$ and $g_{L}^{B\bar{B}^*}\mu_{B\bar{B}^*}$.

We may expect that Eqs. (\ref{DDstar}) and (\ref{BBstar}) could reduce to the same equation in the heavy quark limit. Thus, we assume that with a finite heavy quark mass, this assumption still holds. This requirement leads to
\begin{eqnarray}
g_{L}^{D\bar{D}^*}\mu_{D\bar{D}^*}=g_{L}^{B\bar{B}^*}\mu_{B\bar{B}^*},
\end{eqnarray}
or equivalently,
\begin{eqnarray}
\frac{g_{L}^{D\bar{D}^*}}{g_{L}^{B\bar{B}^*}}=\frac{\mu_{B\bar{B}^*}}{\mu_{D\bar{D}^*}},\label{LEC relation}
\end{eqnarray}
i.e., the LECs of $D\bar{D}^*$ system and $B\bar{B}^*$ system are inversely proportional to their reduced masses. This approximation is roughly consistent with the relations suggested by Refs. \cite{AlFiky:2005jd,Baru:2018qkb}.

We use Eq. (\ref{LEC relation}) to estimate the LECs of the considered systems. Explicitly, the following relations are used in our calculation
\begin{eqnarray}
g_L^{D\bar{D}^*}&=&\frac{\mu_{D^*\bar{D}^*}}{\mu_{D\bar{D}^*}}g_L^{D^*\bar{D}^*}=\frac{\mu_{B^{(*)}\bar{B}^*}}{\mu_{D\bar{D}^*}}g_L^{B^{(*)}\bar{B}^*},\label{DDBB}\\
g_L^{D\bar{D}_s^*}&=&\frac{\mu_{D^*\bar{D}_s^*}}{\mu_{D\bar{D}_s^*}}g_L^{D^*\bar{D}_s^*}=\frac{\mu_{B^{(*)}\bar{B}_s^*}}{\mu_{D\bar{D}_s^*}}g_L^{B^{(*)}\bar{B}_s^*},\label{DDSBBS}
\end{eqnarray}
{ where $g_L^{D^{(*)}\bar{D}_{(s)}^{*}}$ ($g_L^{B^{(*)}\bar{B}_{(s)}^{*}}$) and $\mu_{D^{(*)}\bar{D}_{(s)}^*}$ ($\mu_{B^{(*)}\bar{B}_{(s)}^*}$) are the LEC} and reduced mass of the $D^{(*)}\bar{D}_{(s)}^*$ ($B^{(*)}\bar{B}_{(s)}^*$) and $D^{(*)}\bar{D}_{(s)}^*$ ($B^{(*)}\bar{B}_{(s)}^*$) states, respectively.

\begin{figure*}[htbp]
    \centering
    \includegraphics[width=1.0\linewidth]{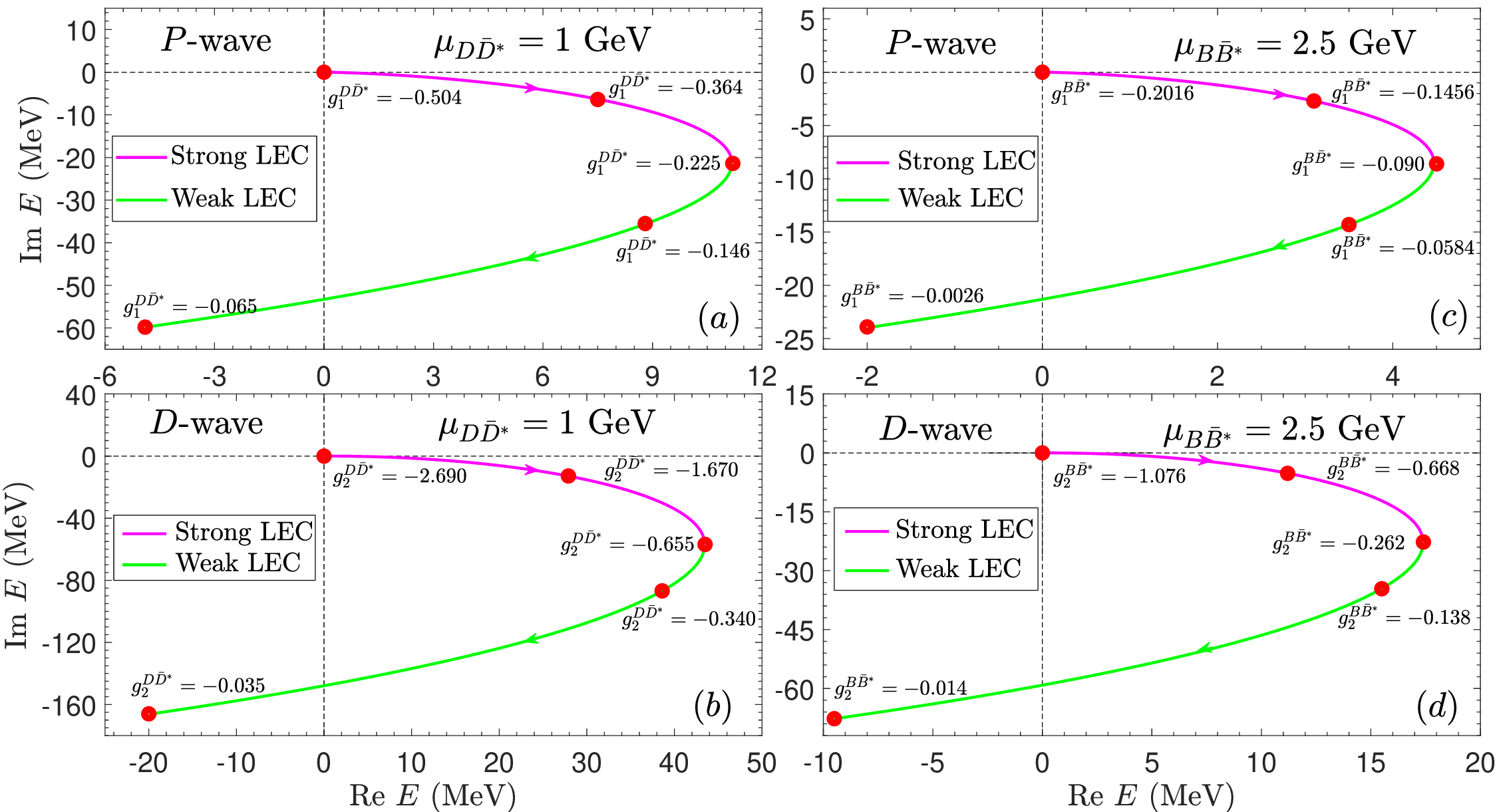}
    \caption{The $P$-wave pole trajectories with $\mu_{D\bar{D}^*}=1$ GeV (a) and $\mu_{B\bar{B}^*}=2.5$ GeV (c) in the resonance region, respectively, and the $D$-wave pole trajectories with $\mu^{D\bar{D}^*}=1$ GeV (b) and $\mu^{B\bar{B}^*}=2.5$ GeV (d) in the resonance region, respectively. The arrows illustrate the directions of decreasing the attractive forces. In each figure, we divide the LEC region into a strong LEC region and a weak LEC region, and we list the LEC values of five resonance poles labeled with red circles. The LECs $g_1^{D\bar{D}^*}$/$g_1^{B\bar{B}^*}$ and $g_2^{D\bar{D}^*}$/$g_2^{B\bar{B}^*}$ are in units of 10$^3$ GeV$^{-4}$ and 10$^{3}$ GeV$^{-6}$, respectively.}
    \label{DBRPD}
\end{figure*}

To obtain a more concrete understanding of the approximation in Eq. (\ref{LEC relation}), in Fig. \ref{DBRPD}, we set $\mu_{D\bar{D}^*}=1$ GeV (close to the reduced mass of $D\bar{D}^{*}$) and $\mu_{B\bar{B}^*}=2.5$ GeV (close to the reduced mass of $B\bar{B}^{*}$), and illustrate the $P$-wave and $D$-wave pole trajectories in the resonance region at $E$-plane, { in which} the arrows denote the decreasing of attractive forces. In Figs. \ref{DBRPD} (a) and (c), we label five $P$-wave $D\bar{D}^*$ and $B\bar{B}^*$ resonance poles and list their $g_1^{D\bar{D}^*}$ and $g_1^{B\bar{B}^*}$ values, respectively. Similarly, in Figs. \ref{DBRPD} (b) and (d), we label five $D$-wave $D\bar{D}^*$ and $B\bar{B}^*$ resonance poles and list their $g_2^{D\bar{D}^*}$ and $g_2^{B\bar{B}^*}$ values, respectively. 
As demonstrated by the numerical values of $g_L^{D\bar{D}^*}$ and $g_L^{B\bar{B}^*}$ ($L=1$, 2) in Fig. \ref{DBRPD}, a $D\bar{D}^*$ pole solution obtained at $g_L^{D\bar{D}^*}$ corresponds to a $B\bar{B}^*$ pole solution obtained at
\begin{eqnarray}
g_L^{B\bar{B}^*}=g_L^{D\bar{D}^*}\frac{\mu_{D\bar{D}^*}}{\mu_{B\bar{B}^*}}=0.4g_{L}^{D\bar{D}^*}.
\end{eqnarray}
Indeed, from Eqs. (\ref{DDstar}-\ref{BBstar}), these two poles have the same complex momentum $k_0$, but corresponds to different energies, i.e.,
\begin{eqnarray}
k_0&=&\sqrt{2\mu_{D\bar{D}^*}\left(E^R_{D\bar{D}^*}-i\frac{\Gamma_{D\bar{D}^*}}{2}\right)},\\
k_0&=&\sqrt{2\mu_{B\bar{B}^*}\left(E^R_{B\bar{B}^*}-i\frac{\Gamma_{B\bar{B}^*}}{2}\right)}.
\end{eqnarray}
Note that the ($E^R_{D\bar{D}^*}$, $-i\frac{\Gamma_{D\bar{D}^*}}{2}$) and ($E^R_{B\bar{B}^*}$, $-i\frac{\Gamma_{B\bar{B}^*}}{2}$) are the pole positions of the $D\bar{D}^*$ state and $B\bar{B}^*$ state at energy plane, respectively. Consequently, the pole positions between the $D\bar{D}^*$ state and $B\bar{B}^*$ state have the relations
\begin{eqnarray}
\mu_{D\bar{D}^*} E^R_{D\bar{D}^*}&=&\mu_{B\bar{B}^*} E^R_{B\bar{B}^*},\label{ER}\\
\mu_{D\bar{D}^*}\Gamma_{D\bar{D}^*}&=&\mu_{B\bar{B}^*}\Gamma_{B\bar{B}^*}.\label{Gamma}
\end{eqnarray}
Thus, with $\mu_{D\bar{D}^*}<\mu_{B\bar{B}^*}$, the mass of $B\bar{B}^*$ state { lies} closer to its corresponding threshold than that of the $D\bar{D}^*$ state, and the width of $B\bar{B}^*$ state is narrower than that of $D\bar{D}^*$ state. Similar relations also exist between the $D^*\bar{D}^*$ resonance and $B^*\bar{B}^*$ resonance. Here, we would like to give a reminder that the experimental resonance parameters of the $Z_c(3900)$/$Z_{b}(10610)$ and $Z_{c}(4020)$/$Z_{b}(10650)$ roughly satisfy { the relation in } Eqs. (\ref{ER}-\ref{Gamma}).

Another important phenomenon in Figs. \ref{DBRPD} (a)-(d) is that for the $P$-wave or $D$-wave resonance above the threshold, one resonance mass corresponds two different widths, depending on the selected LECs. As illustrated in Figs. \ref{DBRPD} (a)-(d), a strong LEC leads to a resonance state with narrow width, while a weak LEC leads to a resonance state with broad width. This fact is crucial for our model to clarify the broad widths of $Z_{cs}(4000)$ and $Z_{cs}(4220)$ states if we assume that these two states are $D$-wave $D\bar{D}_s^*$ and $D^*\bar{D}_s^*$ resonances, respectively.



\section{Numerical results}\label{results}
With the above preparations, now we present our results by adopting the physical reduced masses of $D^{(*)}\bar{D}_{(s)}^*$ and $B^{(*)}\bar{B}^*_{(s)}$ systems. In our calculation, we use the isospin averaged masses of $D^{(*)}_{(s)}$ and $B^{(*)}_{(s)}$ mesons.


{For the $S$-wave $|^1_1[D\bar{D}^*]^S_1,-\rangle$ channel, if the interaction between the $D$ and $\bar{D}^*$ allows the existence of a bound/virtual state, it should lie below the $D\bar{D}^*$ threshold. As discussed in Sec. \ref{Pole trajectories}, for an $S$-wave bound/virtual state pole, there is a one-to-one correspondence between the obtained mass and LEC $g_0^{D\bar{D}^*}$, i.e., the LEC $g_0^{D\bar{D}^*}$ is solely constrained by the input mass of $|^1_1[D\bar{D}^*]^S_1,-\rangle$ state. However, there is no reliable mass input to the $|^1_1[D\bar{D}^*]^S_1,-\rangle$ state, so we omit the calculation and discussion of $S$-wave $D^{(*)}\bar{D}_{(s)}^{(*)}$/$B^{(*)}\bar{B}_{(s)}^{(*)}$ states in this work.}

For the $P$-wave $|^1_0[D\bar{D}^*]^P_1,-\rangle$ state, $D$-wave $|^1_1[D\bar{D}^*]_1^D,-\rangle$ state, and its SU(3)$_{\text{f}}$ partner $|^1_1[D\bar{D}_s^*]_1^D\rangle$ state, we omit the possibilities that they can form bound states, since it should be very difficult for these higher partial states to overcome the central barrier repulsive forces and to form bound states. Instead, we mainly focus on the possibilities that they can form resonances. The LECs $g_1^{D\bar{D}^*}$, $g_2^{D\bar{D}^*}$, and $g_2^{D\bar{D}_s^*}$ can be constrained by both the masses and widths from $|^1_0[D\bar{D}^*]^P_1,-\rangle$, $|^1_1[D\bar{D}^*]_1^D,-\rangle$, and $|^1_1[D\bar{D}_s^*]_1^D\rangle$ candidates. 
\subsection{$P$-wave states}
Here, we use the mass of experimentally observed $G(3900)$ as input to fix the parameter $g_1^{D\bar{D}^*}$ for the $P$-wave $|^1_0[D\bar{D}^*]^P_1,-\rangle$ state. The experimental central value of $M_{G(3900)}$ is 3872.5 MeV \cite{BESIII:2024ths}, below the $D\bar{D}^*$ threshold, and the experimental central value of $\Gamma_{G(3900)}$ is 179.7 MeV. As plotted in Fig. \ref{DBRPD} (a), as a broad $P$-wave state, its LEC $g_1^{D\bar{D}^*}$ should lie in the weak LEC region. In this region, varying the central value of $M_{G(3900)}$ around the $D\bar{D}^*$ threshold by a few MeV will not change the magnitude of its width. Thus, we do not consider the isospin breaking effect introduced from the mass difference between $D^+\bar{D}^{*-}$ and $D^0\bar{D}^{*0}$ components.

{ With the experimental mass of $G(3900)$, we solve the LEC $g_1^{D\bar{D}^*}$ and width of the $|^1_0[D\bar{D}^*]^P_1,-\rangle$ state from Eqs. (\ref{ReRsolution}-\ref{ImRsolution}). We further give their uncertainties by considering the experimental statistical and systematic uncertainties from its mass. Then we calculate the LECs for $|^1_0[D^*\bar{D}^*]_1^P,-\rangle$, $|^1_0[B\bar{B}^*]_1^P,-\rangle$, and $|^1_0[B^*\bar{B}^*]_1^P,-\rangle$ states from Eq. (\ref{DDBB}), and further use these obtained LECs to estimate their corresponding masses and widths.}

\begin{table*}[htbp]
\renewcommand\arraystretch{1.5}
\caption{The results of $P$-wave $D^{(*)}\bar{D}^*$ and $B^{(*)}\bar{B}^*$ states with quantum numbers $I(J^{PC})=0(1^{--})$, and they are obtained from model I and model II. We use the left superscripts ``$\dagger$" to denote our inputs. We only list the LEC $g_1^{D\bar{D}^*}$ for the $|^1_0[D\bar{D}^*]_1^P,-\rangle$ channel, and the LECs $g_1^{D^*\bar{D}^*}$, $g_1^{B\bar{B}^*}$, and $g_1^{B^*\bar{B}^*}$ for the $|^1_0[D^*\bar{D}^*]_1^P,-\rangle$, $|^1_0[B\bar{B}^*]_1^P,-\rangle$, and $|^1_0[B^*\bar{B}^*]_1^P,-\rangle$ channels can be directly obtained from Eq. (\ref{DDBB}).}
\setlength{\tabcolsep}{1mm}{
\begin{tabular}{c|cc|cccccccccccc}
\toprule[0.8pt]
&\multicolumn{2}{c|}{Model I ($g_1^{D\bar{D}^*}=-72.7^{+31.5}_{-131.5}$ GeV$^{-4}$)}&\multicolumn{2}{c}{Model II ($g_1^{D\bar{D}^*}=-58.6^{+33.8}_{-270.4}$ GeV$^{-4}$ )} \\
\hline
State       &Mass (MeV)&Width  (MeV)&Mass (MeV)&Width (MeV)\\
\hline
$|^1_0[D\bar{D}^*]_1^P,-\rangle$&${}^{\dagger}3872.5\pm14.5$&$119.0^{+30.8}_{-65.9}$&
${}^{\dagger}3872.5\pm14.5$&$125.4_{-85.4}^{+51.6}$\\
$|^1_0[D^*\bar{D}^*]_1^P,-\rangle$&$4014.0\pm14.0$&$114.8_{-63.7}^{+27.6}$&
$4014.0^{+12.2}_{-14.1}$&$120.8_{-82.3}^{+50.7}$\\
$|^1_0[B\bar{B}^*]_1^P,-\rangle$&$10603.1\pm5.3$&$43.4^{+11.3}_{-24.0}$&
$10603.1^{+4.6}_{-5.3}$&$45.8_{-31.2}^{+19.2}$\\
$|^1_0[B^*\bar{B}^*]_1^P,-\rangle$&$10648.3\pm5.3$&$43.2_{-23.9}^{+11.2}$&
$10648.3^{+4.6}_{-5.3}$&$45.6_{-31.1}^{+19.0}$\\
\bottomrule[0.8pt]
\end{tabular}}\label{P-wave}
\end{table*}

In Table \ref{P-wave}, with the form factors introduced from model I and model II, we present our results for the { $P$-wave} $D^{(*)}\bar{D}^*$ and $B^{(*)}\bar{B}^*$ states with quantum numbers $I(J^{PC})=0(1^{--})$. As shown in Table \ref{P-wave}, for each state, the masses and widths obtained from model I and model II are comparable with each other. { Thus, for the $P$-wave resonance states, it will not change our results qualitatively by choosing different form factors, and then we can mainly focus on our results obtained from model I.}

From the model I, the obtained width of $G(3900)$ is about $120$ MeV, consistent with the magnitude of the observed $G(3900)$ \cite{BESIII:2024ths}.
The mass and width of the HQSS partner of $G(3900)$ are obtained as 4014.0 MeV and 114.8 MeV, respectively. The mass of this state is close to the well-established $\psi(4040)$ \cite{ParticleDataGroup:2024cfk}, here we want to emphasis that the interpretation of $\psi(4040)$ as a single resonance is still unclear since this state is close to the $D^*\bar{D}^*$ threshold. Thus, looking for the HQSS partner of $G(3900)$ would be an important test to the $P$-wave resonance nature of $G(3900)$. 

We also predict two $P$-wave $B\bar{B}^*$ and $B^*\bar{B}^*$ resonance states with $J^P=1^-$, and their masses are close to the $B\bar{B}^*$ and $B^*\bar{B}^*$ thresholds, respectively. Their widths are all around 40 MeV and comparable to each other. We notice that these two $P$-wave resonance states have also been predicted within the framework of OBE model \cite{Wang:2025kpm,Wang:2025jec}, but with narrower widths. The differences between the widths of $|^1_0[B\bar{B}^*]_1^P,-\rangle$ state and $|^1_0[B^*\bar{B}^*]_1^P,-\rangle$ state in our EFT and the OBE model may mainly due to {the choice of different form factors. It should be noted } that comparing to the form factors adopted in Refs. \cite{Wang:2025kpm,Wang:2025jec}, the form factors we adopted in model I and model II have stronger suppressions on the higher momenta.

\subsection{$D$-wave states}
To estimate the LEC $g_2^{D\bar{D}^*}$ for $D$-wave $D\bar{D}^*$ state with quantum numbers $IJ^{PC}=1(1^{+-})$, { we assume that $Z_c(3900)$ is the candidate of $|^1_1[D\bar{D}^*]^D_1,-\rangle$ resonance state and the mass of $|^1_1[D\bar{D}^*]^D_1,-\rangle$ state is 15 MeV above the $D\bar{D}^*$ threshold with  a uncertainty of $\pm 15$ MeV}. As plotted in Fig. \ref{DBRPD} (b), to obtain a narrow resonance, the LEC should be selected in the strong LEC region, and then we can fix the $g^{D\bar{D}^*}_2$ by the assumed mass of $|^1_1[D\bar{D}^*]^D_1,-\rangle$ state and further  use Eq. (\ref{DDBB}) to estimate the LECs $g_2^{D^*\bar{D}^*}$, $g_2^{B\bar{B}^*}$, $g_2^{B^*\bar{B}^*}$ of $|^1_1[D^*\bar{D}^{*}]^D_1,-\rangle$, $|^1_1[B\bar{B}^*]_1^D,-\rangle$, and $|^1_1[B^*\bar{B}^*]^D_1,-\rangle$ states, respectively.

Similarly, the broad $Z_{cs}(4000)$ could be a $D$-wave $D\bar{D}^*_s$ resonance candidate with quantum number $I(J^P)=\frac{1}{2}(1^+)$, as plotted in Fig. \ref{DBRPD} (b), to obtain a broad resonance, the LEC should be selected in the weak LEC region. We can fix the $g_2^{D\bar{D}_s^*}$ by the experimental mass of $Z_{cs}(4000)$ and use Eq. (\ref{DDSBBS}) to estimate the LECs $g_2^{D^*\bar{D}^*_s}$, $g_2^{B\bar{B}^*_s}$, and $g_2^{B^*\bar{B}_s^*}$ of the $|^1_{\frac{1}{2}}[D^*\bar{D}_s^*]_1^D\rangle$, $|^1_{\frac{1}{2}}[B\bar{B}^*_s]^D_1\rangle$, and $|^1_{\frac{1}{2}}[B^*\bar{B}^*_s]^D_1\rangle$ states, respectively.

\begin{table*}[htbp]
\renewcommand\arraystretch{1.5}
\caption{The results of the $D$-wave $D^{(*)}\bar{D}^*$ ($B^{(*)}\bar{B}^*$) and $D^{(*)}\bar{D}_s^{(*)}$ ($B^{(*)}\bar{B}_s^{(*)}$) states with quantum numbers $I(J^{PC})=1(1^{+-})$ and $I(J^P)=\frac{1}{2}(1^+)$, respectively. They are calculated from model I and model II. We use the left superscript ``$\dagger$" to denote our inputs and list the LECs $g_2^{D\bar{D}^*}$ and $g_2^{D\bar{D}_s^*}$ for the $|^1_1[D\bar{D}^*]_1^D,-\rangle$ and $|^1_{\frac{1}{2}}[D\bar{D}_s^*]_1^D\rangle$ channels, respectively. The LECs $g_2^{D^*\bar{D}^*}$, $g_2^{B\bar{B}^*}$, and $g_2^{B^*\bar{B}^*}$ for the $|^1_1[D^*\bar{D}^*]_1^D,-\rangle$, $|^1_1[B\bar{B}^*]_1^D,-\rangle$, and $|^1_1[B^*\bar{B}^*]_1^D,-\rangle$ channels can be directly obtained from Eq. (\ref{DDBB}), while the LECs $g_2^{D^*\bar{D}_s^*}$, $g_2^{B\bar{B}_s^*}$, and $g_2^{B^*\bar{B}_s^*}$ for the $|^1_{\frac{1}{2}}[D^*\bar{D}_s^*]_1^D\rangle$, $|^1_{\frac{1}{2}}[B\bar{B}_s^*]_1^D\rangle$, and $|^1_{\frac{1}{2}}[B^*\bar{B}^*_s]_1^D\rangle$ channels can be directly obtained from Eq. (\ref{DDSBBS}).}
\begin{tabular}{c|cc|cc|cccccccc}
\toprule[0.8pt]
             &\multicolumn{2}{c|}{Model I ($g^{D\bar{D}^*}_2=-2.274_{-0.503}^{+0.600}$ $10^3$ GeV$^{-6}$)} &\multicolumn{2}{c|}{Model II ($g^{D\bar{D}^*}_2=7.640_{-5.210}^{+5.460}$ $10^3$ GeV$^{-6}$)}&\multicolumn{2}{c}{Exp.}\\
             \hline
             &Mass (MeV)&Width (MeV)                   &Mass (MeV)&Width (MeV)                &Mass (MeV)    &Width (MeV)         & \\
             \hline
$D\bar{D}^*$&    ${}^{\dagger}3890.8\pm15.0$&$6.4_{-6.4}^{+22.7}$                        &${}^{\dagger}3890.8\pm15.0$     &$11.0_{-11.0}^{+51.6}$                      &$3887.1\pm2.6$&$28.4\pm2.6$   \\
$D^*\bar{D}^*$     &$4031.6_{-14.5}^{+14.4}$                       &$6.0_{-6.0}^{+22.1}$     &$4031.6_{-14.5}^{+14.4}$                      &$10.5_{-10.5}^{+49.9}$&$4024.1\pm1.9$&$13\pm5$     \\
$B\bar{B}^*$        &$10609.8\pm5.5$&$2.3_{-2.3}^{+8.3}$     &$10609.8\pm5.5$&$4.0_{-4.0}^{+18.8}$                      &$10607.2\pm2.0$&$18.4\pm2.4$     \\
$B^*\bar{B}^*$      &$10655.0\pm5.5$                     &$2.2_{-2.2}^{+8.4}$     &$10655.0\pm5.5$
&$4.0_{-4.0}^{+18.8}$&$10652.2\pm1.5$&$11.5\pm2.2$     \\
\hline
&\multicolumn{2}{c|}{Model I ($g^{D\bar{D}_s^*}_2=-0.164^{+0.072}_{-0.065}$ 10$^3$GeV$^{-6}$)} & \multicolumn{2}{c|}{Model II ($g_2^{D\bar{D}_s^*}=-0.033_{-0.225}^{+0.029}$ $10^{3}$GeV$^{-6}$)}&\multicolumn{2}{c}{Exp.}&\\
\hline
&Mass (MeV)&Width (MeV)&Mass (MeV)&Width (MeV)&Mass (MeV)&Width (MeV)\\
\hline
$D\bar{D}_s^*$&${}^{\dagger}4003.0^{+7.2}_{-15.2}$&$234.8_{-26.0}^{+41.5}$&${}^{\dagger}4003.0^{+7.2}_{-15.4}$&$294.2_{-131.6}^{+196.6}$&$4003\pm6^{+4}_{-14}$&$131\pm15\pm26$\\
$D^*\bar{D}_s^*$&$4144.7_{-14.5}^{+6.9}$&$226.5_{-25.3}^{+39.9}$&$4144.7_{-14.6}^{+6.9}$&$283.6_{-126.8}^{+189.6}$&$4216\pm24^{+43}_{-30}$&$233\pm52^{+97}_{-73}$ \\
$B\bar{B}^*_s$&$10702.5_{-5.6}^{+2.7}$&$87.2^{+15.4}_{-9.7}$&$10702.5_{-5.6}^{+2.7}$&$109.3_{-48.9}^{+73.0}$\\
$B^*\bar{B}^*_s$&$10749.4_{-5.6}^{+2.7}$&$86.8^{+15.4}_{-9.6}$&$10749.4_{-5.6}^{+2.7}$&$108.8_{-48.7}^{+72.6}$\\
\bottomrule[0.8pt]
\end{tabular}\label{D-wave}
\end{table*}

We present the results of $D$-wave $D^{(*)}\bar{D}^*$/$B^{(*)}\bar{B}^{(*)}$ $I^{G}J^{PC}=1^+(1^{+-})$ states obtained from model I and model II in the upper panel of Table \ref{D-wave}. As shown in Table \ref{D-wave}, the results obtained { from} model I and model II are comparable to each other, so we mainly discuss the results of model I.

By including the uncertainties of our results, the width of the $D$-wave $|^1_1[D\bar{D}^*]^D_1,-\rangle$ state is comparable with that of the observed $Z_c(3900)$. Similarly, the mass and width of its HQSS partner $|^1_1[D^*\bar{D}^{*}]^D_1,-\rangle$ state are comparable with that of the observed $Z_c(4020)$. Our results indicate that the experimentally observed $Z_c(3900)$ and $Z_c(4020)$ could be the $|^1_1[D\bar{D}^*]^D_1,-\rangle$ and $|^1_1[D^*\bar{D}^{*}]^D_1,-\rangle$ resonance candidates, respectively.


{ In our framework, the LECs of the $S$-wave bound/virtual states are solely determined by their masses. Thus, without reliable mass inputs, we can not give reliable predictions to the $S$-wave bound/virtual states.} Nevertheless, our framework also allows the existence of an $S$-wave virtual state by appropriately adjusting the LEC $g_0^{D\bar{D}^*}$. Besides, from Refs. \cite{Lin:2024qcq,Yu:2024sqv,He:2017lhy,Albaladejo:2015lob,Du:2022jjv,Gong:2016hlt}, the $Z_c(3900)$ can be interpreted as an $S$-wave virtual state below the $D\bar{D}^*$ threshold. Thus, a very interesting possibility for the observed $Z_c(3900)$ is that it may contain two states, the $S$-wave $Z_c^V(3900)$ (virtual state) and $D$-wave $Z_c^R(3900)$ (resonance state). The $Z^V(3900)$ and $Z^R(3900)$ lie below and above the $D\bar{D}^*$ threshold by a few MeV, respectively. Due to the molecular configurations of these two states, they lie very close to each other. This two-peak hypothesis provides a new insight to understand the underlying structure of the observed $Z_c(3900)$.

The results of $D$-wave $B\bar{B}^*$ and $B^*\bar{B}^*$ states with quantum numbers $IJ^{PC}=1(1^{+-})$ are also given in Table \ref{D-wave}. By considering the uncertainties from the assumed mass of the $|^1_1[D\bar{D}^*]^D_1,-\rangle$ resonance, we find that our results are consistent with the measured resonance parameters of $Z_b(10610)$ and $Z_b(10650)$ \cite{ParticleDataGroup:2024cfk}.



We present the results of $D$-wave $D^{(*)}\bar{D}_s^*$/$B^{(*)}\bar{B}_s^{(*)}$ $IJ^{P}=\frac{1}{2}(1^{+})$ states from model I and model II in the lower panel of Table \ref{D-wave}. { The magnitude of $\Gamma_{|^1_{\frac{1}{2}}[D\bar{D}_s]_1^D\rangle}$ is comparable with that of the $Z_{cs}(4000)$ reported by the LHCb collaboration} \cite{LHCb:2023hxg,LHCb:2021uow}. Note that this broad state is obtained by selecting a weak LEC, which is different from the narrow $|^1_1[D\bar{D}^*]_1^D,-\rangle$ state obtained by selecting a strong LEC. The different interactions between the $|^1_1[D\bar{D}^*]_1^D,-\rangle$ state and its SU(3)$\text{f}$ partner $|^1_{\frac{1}{2}}[D\bar{D}_s]_1^D\rangle$ state are consistent with the general expectation. For example, from the OBE model, the $|^1_{\frac{1}{2}}[D\bar{D}_s^*]_1^D\rangle$ state { has fewer exchanged light mesons comparing to that} of the $|^1_1[D\bar{D}^*]_1^D\rangle$ state, so that the $|^1_{\frac{1}{2}}[D\bar{D}_s^*]_1^D\rangle$ state may have a weaker attractive force. Thus, the weak LEC of the $|^1_{\frac{1}{2}}[D\bar{D}_s^*]_1^D\rangle$ state leads this state to be a good candidate of $D$-wave $D\bar{D}_s^*$ resonance. Besides, we also predict its HQSS partner composed of $D^*$ and $\bar{D}^*_s$ mesons. As presented in Table \ref{D-wave}, this broad state may correspond to the $Z_{cs}(4220)$ \cite{LHCb:2023hxg,LHCb:2021uow} observed in the $J/\Psi K$ final state. Similarly, with a weak LEC $g_2^{D^*\bar{D}^*_s}$, the $Z_{cs}(4220)$ is also a good candidate of $D$-wave $D^*\bar{D}_s^*$ resonance.

In Table \ref{D-wave}, we also predict the possible $D$-wave $Z_{bs}$ states in the $|^1_{\frac{1}{2}}[B\bar{B}^{*}_s]_1^D\rangle$ and $|^1_{\frac{1}{2}}[B^*\bar{B}^*_s]^D_1\rangle$ channel. As demonstrated in Eqs. (\ref{ER}-\ref{Gamma}), comparing to the $Z_{cs}(4000)$ and $Z_{cs}(4220)$  states, these two $D$-wave $Z_{bs}$ lie closer to their correspond thresholds and have narrower widths.
\section{Summary}\label{summary}
In this work, we construct an effective field theory and introduce a solvable model to gain some general features of the $S$-wave/$P$-wave/$D$-wave bound/virtual/resonance states in the $D^{(*)}\bar{D}_{s}^*$ and $B^{(*)}\bar{B}_{(s)}^*$ systems.

In the heavy quark limit, we describe their interactions by analogy with the two-nucleon contact terms. The lowest order contributions to the $S$-wave, $P$-wave, and $D$-wave interactions are from the LO, NLO, and N$^3$LO contact terms, respectively. To reduce the number of LECs, we assume that the emergences of the $S$-wave, $P$-wave, and $D$-wave bound/virtual/resonance states are dominated by the LO, NLO, and N$^3$LO contact terms and describe the $S$-wave, $P$-wave, and $D$-wave potentials by only considering their lowest order contributions.

We present the typical pole trajectories for the $S$-wave, $P$-wave, and $D$-wave contact potentials. We show that with strong enough attractive forces, the $S$-wave, $P$-wave, and $D$-wave contact terms all can form bound states. As the attractive forces decrease, the $S$-wave bound states become virtual states, while the $P$-wave and $D$-wave bound states become resonances.

{Since we are mainly interested in the $|^1_1[D\bar{D}^*]^S_1,-\rangle$, $|^1_0[D\bar{D}^*]^P_1,-\rangle$, $|^1_1[D\bar{D}^*]^D_1,-\rangle$, and $|^1_{\frac{1}{2}}[D\bar{D}_s^*]_1^D\rangle$ states, so four LECs $g_0^{D\bar{D}^*}$, $g_1^{D\bar{D}^*}$, $g_2^{D^*\bar{D}^*}$, and $g_2^{D\bar{D}_s^*}$ are introduced, and then we introduce the approximations in Eqs. (\ref{DDBB}-\ref{DDSBBS}) to relate these four LECs to the $D^*\bar{D}_{(s)}^*$, $B^{(*)}\bar{B}_{(s)}^{*}$ states.}
We omit the possibilities that the $P$-wave and $D$-wave contact potentials may form bound states since it should be difficult for the $P$-wave/$D$-wave $D^*\bar{D}_{(s)}^*$, $B^{(*)}\bar{B}_{(s)}^{*}$ states to overcome the repulsive central barrier forces and then become bound states with strong attractive forces. Instead, we mainly focus on the possibilities that they can form resonance states. Besides, for the $S$-wave bound/virtual states, since we do not have reliable inputs, we omit the calculations for the $S$-wave $D^*\bar{D}_{(s)}^*$ and $B^{(*)}\bar{B}_{(s)}^{*}$ states.

Here, we want to emphasis that our results are based on the assumptions of the $P$-wave/$D$-wave resonance natures of the observed $Z_c$ and $Z_{cs}$ states. More serious calculations are still needed to check if their LECs obtained in our EFT framework could be realized physically.

The resonance parameters for the $P$-wave/$D$-wave $D^*\bar{D}_{(s)}^*$ and $B^{(*)}\bar{B}_{(s)}^{*}$ states obtained from our model are roughly consistent with the measured resonance parameters of $G(3900)$, $Z_c(3900)$/$Z_c(4020)$, $Z_b(10610)$/$Z_b(10650)$, and $Z_{cs}(4000)$/$Z_{cs}(4220)$ states, indicating a unified $D$-wave molecular picture to { these isovector charmoniumlike states}. Especially, we predict a $P$-wave $D^*\bar{D}^*$ state, which is the HQSS partner of $G(3900)$. We propose a two-peak hypothesis to interpret the observed $Z_c(3900)$ state, i.e, this state may consist of { an $S$-wave $Z_c^V(3900)$ virtual state and a $D$-wave $Z_c^R(3900)$ resonance state}. Besides, we suggest that due to the weak LEC couplings, the $Z_{cs}(4000)$ and $Z_{cs}(4220)$ are good candidates of $D$-wave resonances. We hope that some of our predictions can be tested by the BESIII, BELLE II, and LHCb collaborations in the future.


\section*{Acknowledgments}
This work is supported by the National Natural Science Foundation of China under Grants No. 12305090, No. 12405088. K. Chen is also supported by the Start-up Funds of Northwest University. J. Z. Wang is also supported by the Start-up Funds of Chongqing University.

\end{document}